# Detection of Information leakage in cloud

MANSAF ALAM , SHUCHI SETHI
Computer Science
Jamia Millia Islamia
New Delhi
INDIA
malam@jmi.ac.in , shuchi.sethi@yahoo.com

*Abstract:* - Recent research shows that colluded malware in different VMs sharing a single physical host may use a resource as a channel to leak critical information. Covert channels employ time or storage characteristics to transmit confidential information to attackers leaving no trail. These channels were not meant for communication and hence control mechanisms do not exist. This means these remain undetected by traditional security measures employed in firewalls etc in a network. The comprehensive survey to address the issue highlights that accurate methods for fast detection in cloud are very expensive in terms of storage and processing. The proposed framework builds signature by extracting features which accurately classify the regular from covert traffic in cloud and estimates difference in distribution of data under analysis by means of scores. It then adds context to the signature and finally using machine learning (Support Vector Machines), a model is built and trained for deploying in cloud. The results show that the framework proposed is high in accuracy while being low cost and robust as it is tested after adding noise which is likely to exist in public cloud environments.

## 1 Introduction

Virtualization is the key enabling technology in cloud computing. On one hand coresidency allows services to be offered at low cost while on the other coresidency with other tenants pose a security threat which is a major barrier to mass adoption of cloud. Platform sharing invariably leads to a new set of emerging threats. Covert channels in shared hardware enable attackers to exfiltrate sensitive data. This study has gained momentum after research by Ritenpart et al.[15] demonstrated the ability of mapping internal cloud infrastructure and estimating the likely placement of target VM and instantiate new VMs until one of them resides with target VM. The secure isolation is broken by covert channels in a virtual environment by leaking data from a program with higher privileges to the ones with lower level of clearance. In cloud this can have serious repercussions in which security of data relies on strong isolation assumption.

There are various categories of covert channels for leaking information like paging rate, temperature based covert channels. But most channels are noisy with high error rates. Certain channels like billing of used resources may conceal secret information.

There are two main criteria for categorization of covert channels. One is on the basis of storage and another one is on the basis of time. Storage based covert channels embeds bits in packets for receiver to directly read. They have higher bandwidth but are relatively easier to detect. In time based channels sender exploit time intervals of various resources to transmit the bits. Time based covert channels engage different resources for sending bits while receiver observes and extracts information from time charateristics. Two main strategies are usually employed to counter time channel attacks. One is fuzzy time where the accuracy of the clock is disrupted so that common time reference of sender(a trojan) and receiver(a spy) is also disrupted. Another way is to use lattice scheduling for processor scheduling to avoid sender and receiver processes be able to run at the same time. With the use of self clocking techniques, popular in network transmissions, time disruption obstacles can be overcome.

It is argued that multitenant systems can be made more secure by identifying possible attacks in the cloud architecture and enabling a mechanism for detection of such attacks. At a high level, this premise is supported by proposed framework for covert channel based anomaly detection in cloud which attributes its origin to multitenancy in cloud. There have been research efforts for detection of covert channel attacks, which is an active research area. Various methods have been proposed for the detection of covert timing channels. Out of various approaches proposed in literature for mitigation and detection,only a few suitable are found suitable for cloud environment and certain techniques of detection which provide high accuracy are very expensive in terms of processing time and storage.

The limitations that exists with current approaches are that they are expensive in terms of resource, punish innocous traffic and have their doubts of practicality due to legal concerns. Also most approaches target very specific single attack which limits its use due to lack of generalizability. In particular, the paper encompasses wide variety of time based covert channel attacks. We have proposed an efficient and effective framework which has balanced learning methodology and use of context enables raise alerts on anomalous use of computing resources leading to enhanced accuracy in detection. In literature survey authors have not found use of context as a parameter in covert channel detection.

Our Contributions can be summarized as follows:
1. Design of an efficient low cost framework for covert channel attack detection in cloud.

2. To enhance the accuracy in detection context is introduced.

3. Exploration of implementation feasibility in distributed systems.

4. Balanced learning approach to enable equal emphasis on false positives and false negatives.

Remainder of this manuscript is organized into various sections. Section 2 gives comprehensive survey of related research work done so far in the field of covert channel detection. Further in section 3, we discuss threat model for our framework, which explains the scope of threats and entities that are part of framework, covered in our research. In next section, we explain how the co-residence is abused and exploited in multitenant architecture. This section also proves the theorem on existence of covert channels in multitenant architecture. In section 5, we explain our framework and validate it with use of experiments and results. Finally manuscript is concluded with a brief summary and future directions.

## 2. Related Work

The need for cloud to be secure has lead to requirement of low cost framework for anomaly detection in cloud infrastructure. Covert channel attacks are such anomalies that are difficult to detect. A number of research papers have investigated the serious concern of Time Based Covert Channel Attacks in cloud environment(TBCA). A few are based on shape and regularity tests and measuring entropy of the data to determine the data is legitimate[3,11].

In [1] authors introduce a statistic called Minimal Requisite Fidelity that can result in disruption of communication in structure carriers and decrease the capacity/speed of the channel whose design allows steganographic analysis of the traffic. Most common approach for network analysis is "the wardens" approach that can detect such communication. There are two types of warden, one is active warden in which the network packets are intercepted and manipulated in the manner that all covert bits would be lost if any but this strategy is costly and manipulates information of innocuous traffic as well. In passive warden approach detection is done by capturing network traffic and only malicious traffic is altered. In certain other approaches [5,7,8,10] data is tested to see if it carries a secret message.

In [16] a solution to such attack is proposed by capturing all atomic instructions and replacing them as Trap to hypervisor. When the hypervisor intercepts this trap, it pauses the guest's vCPUs and executes the instruction without asserting the memory bus lock. Since the guest's vCPUs are paused, no other memory operations can be performed by this guest. But tackling situation in the manner has a limitation of slowing the system down which can have unwanted repercussions .

In [4] researchers have implemented an observer that uses a secure VM to compare traffic between a vulnerable VM and secure VM. This method can be used in real time. But crux here is maintaining a secure copy of every VM which itself becomes its limitation. The framework in [20] offers a flexible system for detection by classifying their distributions as markovian when attack is being launched which attributes to the fact that time intervals placed during attack will be regular and authors have used it along with Bayesian training .This methodology has a prerequisite of strong interdependence assumption, which makes it suitable for certain type of covert channels .

In cloud dynamicity of the system cause error rates to go high requiring new thresholds. Cabuk et al [10] have designed robust time based covert channel and proposed regularity(correlation in data) tests for detection of covert channels. With study of shape(mean ,variance, distribution) applying tests like Kolmogorov-Smirnov authors have claimed that it is possible to detect most covert channels. They have designed two metrics for packet arrival interval, one being regularity measure as mentioned in [20] above and another one being the regularity score which depicts the regularity of traffic with stream. In [11] researchers proposed conditional entropy based approach for covert channel detection. The authors have tested the time stamp on four metrics that are Kolmogorov-Smirnov test score, regularity score , entropy and corrected conditional entropy.

**Table1: Parameters for detection**

| Methods | Parameters | | | | |
|---|---|---|---|---|---|
| | *Shape of traffic* | *Regularity of traffic* | *Distribution of traffic* | *Context* | *Inter arrival time* |
| *Cabuk et al.[10]* | | ✓ | | | ✓ |
| *Gianvecchio et al.[11]* | ✓ | ✓ | ✓ | | |
| *Pradhumna et al.[9]* | ✓ | ✓ | ✓ | | |
| *Our method* | | | ✓ | ✓ | ✓ |

Similarly Berk el al[7] have proposed that probability of traffic being covert is a function of mean delay and total number of packets. The function is given by:

$Pcc = (1-Xmean)/Xtotal$ .

In [23] authors propose a scheme of Homealone that can detect any unusual cache activity when friendly VMs are silenced for a certain period. Though the original approach can be modified to enable detection , but this approach has a high overhead as suggested by researchers in [21].

In [24] Ruby et al. have demonstrated that why and how do covert channel attacks exist in current systems. Authors have proposed theorem and proved pertaining to conditions that must exist for covert channel to exist. The theorem states that if sender is able to invoke change(s) in the visible space of the receiver or sender is able to change when an object is updated relative to the observation made by the receiver, a covert channel may exist.

Our work is related to [9] as they have proposed SVM based classifier, we use the same classifier because of its advantages in terms of accuracy and performance for low dimensional dataset. While authors have used filtering of traffic captured by wireshark, we specifically focus on attacks in cloud and hence approach differs significantly. In our work we study conditions in which covert channel can exist in cloud.In [20] authors have proposed a hybrid approach for cloud based covert channel detection, where two level filtering of traffic happens to obtain thresholds. Assumption of markovian distribution is made in case of covert traffic and then training is based on bayesian classifier. The work is closely related to ours because one of the attack is replicated in similar manner and parameter of time interval is incorporated in our framework.

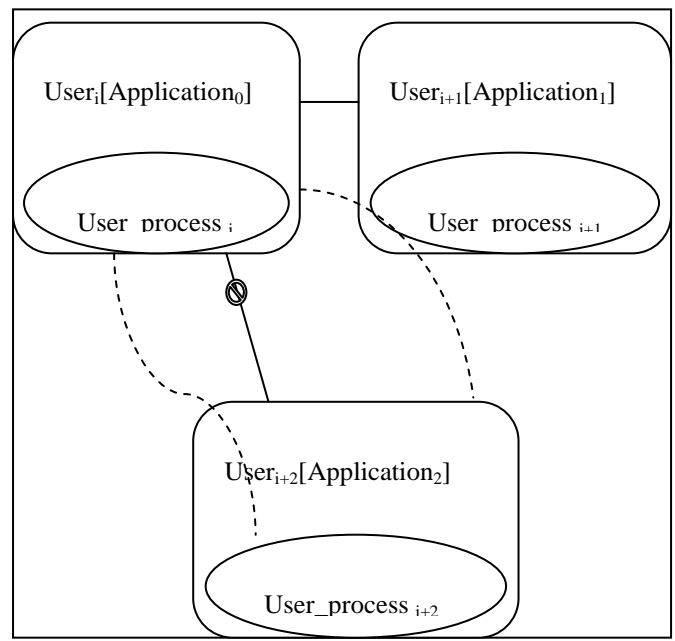

**Figure1:Threat model showing covert communication between $process_i$ and $process_{i+2}$**

Past research in this area has focused less on cloud features of scalability,dynamism and distributed processing. Our research is motivated by the fact that while significant research has been done in detecting TCBA, no approach exists which takes care of context and balanced learning of parameters so that false negatives have equal priority as false positives. These parameters in our approach are important in increasing accuracy and hence reliability.

## 3. Framework Proposed for Detection

This section discusses in detail the proposed framework and methods that lead to accomplishment of objective of detection of covert channels in cloud environment.

### 3.1 System Threat Model

The two popular categories of inter VM covert channel are: if the VMs are on same platform and the other one is when the VMs are on separate platform. The latter category consists of network based covert channels using TCP/IP header information like ISNs which hide confidential information and are randomly generated by the sender. Though such attacks are hard to detect but novel approaches using chaos theory have been successfully applied to detect such covert channels and such attacks are not new to cloud environment. The inter VM attacks on same platform are specific to cloud and bypass methods deployed for network. The threat model includes the attackers which have access to same VM as the victim .The malicious intentions of the attackers can be bypassing access control and leak confidential data or increase use of shared resource to deny essential services to co resident VMs. In cloud environment mainly two categories of attacks exist:
1) Infrastructure based: The attacks that exploit hardware resources for communication like CPU based, cache based, memory based, disk based.
2) Service based: Attacks launched while distribution of resources and during migration fall under this category.

Figure 1 depicts block representation of the threat model that exist in cloud under Infrastructure based attack category,which is in scope of the paper. In case 1 $process_i$ is communicating with $process_{i+2}$ using hardware resource requests. In case 2 $process_i$ is attacking $process_{i+2}$ using another $process_{i+1}$ of a different user application. These types of attacks are usually take place to choke the resources and deny services to user applications. In our paper we explore in detail case 1 of the threat model which involves leaking of information covertly. Threats that break the isolation layer of VMs by leaking information between VMs which are otherwise prohibited from communicating directly and want to share information illegitimately are part of our model.

### 3.2 Methodology

#### 3.2.1 Exploiting Co-residence

There are minimum requirements to set up a covert channel. Theorem 1 stated below was proven by Ruby Lee[24] which pertains to fact that if the visible space (definition 1) exists then the covert channel may exist. But with respect to cloud we have a gap as no theorem has been proven in this regard. In theorem 1, we prove that 1 virtual machine though separated by hypervisor can invoke changes in visible space of another virtual machine.

*Definition 1*: The visible space VS is the set of all objects that the observing application can learn the value of object from its current abstraction level.

*Theorem 1*: If a virtual machine can invoke changes(s) in visible space of other virtual machines,a covert channel may exist.
*Proof*: Let Vn denotes virtual environment which is a set of objects. $Ob_i \varepsilon V_n$ for i=1 to n, for n shared resources inclusive of time slice. Shared resources are abstracted by the layer hypervisor which is providing isolation. Request to a shared resource will be visible to receiver in the form of obtained resource time $T_o$. Thus $T_o$ is dependent on $Ob_i$. Thus the dependency relation can be observed.

Sender→$Ob_j$= access($Ob_j$) ←→$T_o$ →Receiver

This theorem states the necessary condition for covert channel to exist. It does not prove sufficient condition for covert channel to exist but if the given condition exists, the covert channel attack can happen.
There have been research efforts [25] to show in public clouds that co-residency can be detected at very low cost and findings surprisingly show that despite presence of big data centres , it is likely that after a few requests the VM will be co-located with the required VM. This happens due to localized placement policies and thus normalized success rate would be very high. This brings out vulnerability in placement of major cloud service providers Azure,EC2,GCE.

Once co-residency is detected, it would be easy to abuse it with the algorithms used for manipulation of shared resources. Three algorithms are given in next section.

### 3.2.2 TBCA Algorithms

Three TBCA algorithms are studied to test the performance of our framework. A brief description of the algorithms and their implementation in our experiments is provided in this section.

The CPU load based algorithm is based on manipulating CPU load to transmit data secretly where two virtual machines have agreed on time interval for synchronization. To reduce chances in error propagation we have adapted the algorithm(Algorithm 1) in our experiments by keeping large margins(20%).

---

**Algorithm 1 : CPU Load Based Attack**

**Algorithm 1(a)**

**Input** : B=sizeof(File to be leaked), $R_1$= bit 1 ,$R_2$=bit 0, m= threshold of CPU load to marginalize the attack(%), value_high and value_low are system configuration dependent values that can be determined during experimentation.

**Output**: achieved time synchronization

VM1:
      **Begin**

1. While B != EOF
2.    If $R_1$ = = 1
3.      { m = value_high}

//difference between low and high is ~ 20%.
4.     Else
5.      {m = value_low}

//VM1 pause for t sec, causing load to drop.
6.     End If
7.    EndWhile

      **End**

---

**Algorithm 1(b)**

**Input** : value_high and value_low

**Output**: File B received by VM2

VM2:   **Begin**

1.   If $m > \left( \dfrac{Value\_High + Value\_low}{2} \right)$
2.     {$R_1$=$R_1$+1}
3.   Else
4.     { $R_1$=$R_1$+0 }
5.   Endif
   **End**

---

The cache based attacks exploit system cache for the purpose of covert communication. In algorithm adapted from standard cache algorithm, we have ignored certain aspects like scheduling difficulties to keep variables other than time constant. The colluding parties agree upon time intervals for which to keep cache busy and receiver requests for L2cache in those intervals as in most architectures L1 cache is not shared.

---

**Algorithm 2 : Cache Based Attack**

**Algorithm 2(a)**

**Input** : B=Sizeof cache (File to be leaked), $R_1$= bit 1 , $R_2$=bit 0, L= (end-start)/CLK_TCK) \\elapsed time; float start=clock(),Repeatedly read file into cache and flush to obtain estimate average system access latency to obtain $t_1$, Created file F(B);

**Output** : achieved time synchronization

VM1:   **Begin**
1.  While (B)  Do
2.  If (char$_b$ = = $R_1$)
3.  { start=clock()
4.  F(f_open)
5.  end=clock()
6.  }

7. Else

8. {start=clock()

9.  end=clock()  }

10. Endif

11. EndWhile

   **End**

**Algorithm 2(b)**

**Input** : B=Sizeof cache (File to be leaked), $R_1$= bit 1 , $R_2$=bit 0, L= (end-start)/CLK_TCK \\elapsed time; float start=clock(),Repeatedly read file into cache and flush to obtain estimate average system access latency to obtain $t_1$, Created file F(B);

**Output** : File B received by VM2

VM2: **Begin**

1. Start=clock();

2. F(f_open)

3. End=clock();

4. If (L > $t_1$)

5.  { Output=Output+$R_1$}

6. Else

7.  { Output=Output+$R_2$}

8. Endif

   **End**

Memory based attacks model memory bus based attack in which contention of memory bus leads to leakage of bits from sender to receiver. The applications involved in leaking exploit shared resource bus which carries data to and from memory. This means that any request which involves accessing memory will ensure that bus is used. Certain commands like xchg allow automatic locking of memory bus before transaction can be completed while other commands need to be provided with prefix lock to ensure bus is locked before transmission of data between memory and registers.

**Algorithm 3: Memory Based Attack**

**Algorithm 3(a)**

**Input**: N = no of bits to be sent, $R_1$= bit 1 ,$R_2$=bit 0,L= (end-start)/CLK_TCK \\elapsed time ; float start=clock();

**Output** : achieved time synchronization

VM1:    **Begin**

1. While (B)   Do

2.     If (char(B) = = $R_1$)

For t time put memory bus in contended state

3.     Else

For t time leave memory bus in free state

4.     End if

5. Endwhile

   **End**

**Algorithm 3(b)**

**Input**: N = no of bits to be sent, $R_1$= bit 1 ,$R_2$=bit 0,L= (end-start)/CLK_TCK \\elapsed time ; float start=clock();

**Output** : File B received by VM2

VM2:   Begin

1. Start=clock()

2. Xchg mem_address, ax

3. End=clock()

4. If (L > $t_1$)

5. {Output=Output+$R_1$}

6. Else

```
7. { Output=Output+R₂ }
8. Endif
End
```

## 3.3 Framework Overview

In this section, we present an overview of our proposed framework for covert channel based anomaly detection in cloud. Figure 2 shows block diagram representation of system model proposed for detection of covert channel based anomaly in cloud. It consists of 3 primary modules: a captor, signature builder and SVM (support vector machine) based classifier.

The captor contains code for capturing data. The alternative approach could be to place a hook in the hypervisor. But this is applicable only in open source systems. In some cases system logs can also be used for the purpose. In cloud full access to log files might not be allowed as part of policy matters. The second one is the signature builder which is discussed in detail in the following section. In brief parameters resource time interval, resource usage(%) as context, k-s test score are used to build signature.

The value of k-s test score will depend on null hypothesis. If null hypothesis is rejected, score field will record 1 or if it fails to reject then score will be captured 0. A three tuple information for multiple record is passed to classifier $<α,ß,c>$ where α stores time interval of resource obtained c, ß contains k-s test score and c denotes context. After training model can be deployed on live cloud environment for detecting covert channels.

### 3.3.1 Signature Builder Process

To initialize signature building process, we needed to acquire feature points for which we ran cronjob on ubuntu system to capture resources used time, percentage and certain other features obtained with top command and redirected its output to a file to log entries. The file records request of all VMs running in the setup. Relevant fields are filtered for feature extraction. This file currently contains overt traffic for our framework. Then for embedding TBCAs we rotate time interval by 7sec,10sec,20 sec after every 2 minute interval to avoid creating clear patterns for detection framework. The intervals are large so as to make detection harder by creating a low bandwidth channel. The receiver was synchronized to observe the changes in access latency. The features obtained for the purpose of building signature were resource usage time interval of application processes, two sample k-s test score, context variable percentage of resource usage. Data is divided into blocks and for each block k-s test score is calculated and stamped corresponding to each data point in the block. This is done to create matrix input for SVM classifier. Use of k-s test score as parameter will allow higher accuracy than just specifying time interval because in cloud due to scheduling patterns of processes, it is imperative that resource intervals in innocuous traffic will have certain patterns and it will not be completely random. Also additional parameter context will further fortify the framework in cloud as in certain cases patterns of resources will be misleading, so using amount of resource which will be standard for single application creating need for resource, helps maintain accuracy in detection. Besides as part of future work we would like to scale this framework's utility to incorporate anomaly detection. The use of context variable will be required.

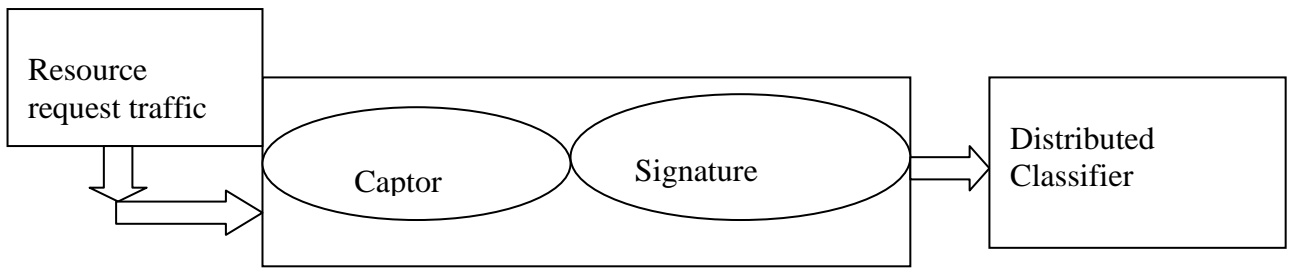

**Figure 2: The Proposed Framework**

We selected a 10,000 samples from each week's readings from experiment run for 4 weeks. These 40,000 data points were divided into chunks of 5000 samples each. The 8 blocks thus formed are used to build signature for the distributed classification. The purpose of large block sizes is to enable faster processing in distributed learning environment. The data set so obtained is re-divided to have smaller block size this time to study impact on detection process. This holds for cases where covert attacks are done just to leak keys and not data file. Detection framework should be able work on small training dataset as cloud is dynamic environment and migration happens frequently. Though for processing and training on large data in less time is made feasible by distributed algorithm but at certain times samples less than a day's log may only be available which on an average for slow speed channel is ~300 samples. Since small sample set doesn't represent the population well, any detection mechanism based on distribution assumption or parametric statistical approach will not perform well. Thus non parametric tests are more suitable as well as use of machine learning algorithms are imperative. Thus we obtain 200 sample block from 40,000 data points which is less than a day's run. The features are then added to dataset by calculating scores for each block and spreading it against each data point in the block to create matrix. Then to ensure balanced learning we take approach of equal data points for positive and negative samples. From each block we segregate positives and negatives and then code written in matlab is run to pick top n/2 samples of positives and top n/2 samples of negatives. If the negative samples are less than n/2, using stamping negative samples were just repeated randomly to obtain n/2 samples where n is size of the block.

### 3.3.2 Proposed Algorithm

In this section we discuss in detail the algorithm we have proposed for implementation of the framework discussed in the last section. The algorithm runs in two phases, initial phase builds signature based on identified parameters while in second phase this signature output becomes input and distributed learning of the algorithm takes place.

---

**ALGORITHM 4 : Kernel Context_Covert Channel Detection**

**Input** :User[m][n] is input file representing resource usage activity of n applications containing m features. User_time (ut) represents m dimensional array containing resource usage time request/allocation of n users. m, n ε N, f is distribution function obtained from sampled values while g is gaussian distribution function, z denotes amount of resource usage.

**Output**: A three tuple signature $<\alpha,\beta,c>$
  Begin
  1. **For i = 1 to n do**
  2. **For j=1 to m do**
  3.     $x_i = ut_{ij}+1 - ut_{ij}$
  4. **EndFor**
  5. **EndFor**
  6. **For i= 1 to k**
  7. $\beta_i = \dfrac{x_i - mean(x_i)}{\sigma x_i}$
  8. $\alpha_i = \sup_{x}|f(x_i) - g(x_i)|$
  9. $c_i = z_i$
  10. **EndFor**
  End

---

**ALGORITHM 5. Distributed Classification**

**Input:** $VM_i$, $<\alpha,\beta,c>$, n is total number of data points and m is number of VMs, sv denotes support vectors

**Output:** Output[n], array of decision variables denoting attack or innocuous traffic.

  Begin
  1. **For** i=1 to m
  2. **For** $VM_i = <\alpha,\beta,c>$
  3.     Result=$sv_i$.alpha$||sv_{i+1}$.alpha
  4.     Output$_n$= Result >0
  5. **Endfor**
  6. **Endfor**
  End

---

The algorithm 4 is proposed is input to algorithm 5. The steps discuss in detail the process to be followed. In step 1 time interval of resource request is calculated from initial and final time stamps of resource requests by a process.

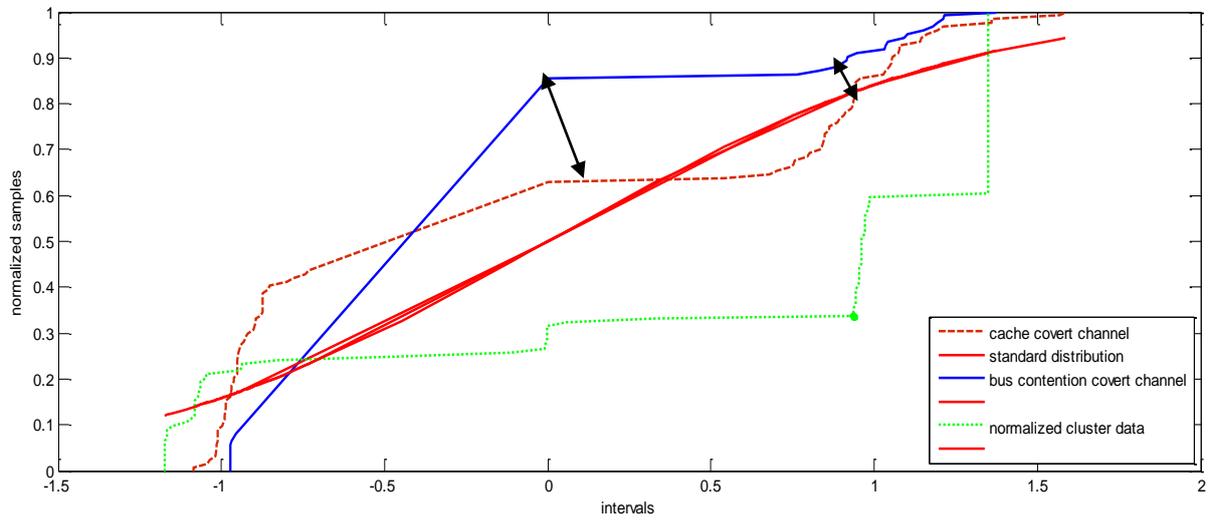

**Figure 3 : graph showing k-s test on overt and covert channels**

In step 3 sample values are obtained after rotating intervals to obtain a vector x[k]. In step 7 using z-score normalization values are obtained in $ß_i$ so as to avoid learning by algorithm from magnitude of values. In step 8, the vector contained in $α_i$ is k-s test score . Vector c contains context variable and is assigned values in step 9. This signature is input to algorithm 5, which distributes the signature to multiple VMs in step 2 such that each VM has equal number of positive and negative training examples :

$$\forall VMi \leftarrow \alpha_i, \beta_i, C_i \quad \text{(process of mapping)}$$

$$sv = \bigcup_{i=1}^{m} sv_i \quad \text{for } sv.\alpha > 0 \quad \text{(process of reducing)}$$

Then output of each VM is recorded and filtered such that positive alphas from the output of the structure are picked in step 4. The so obtained structure after training is now ready to be deployed and tested in cloud.

## 4. Experiments and Discussion

In this section, we present the results obtained by our framework using support vector machine based classifier. Results presented pertain to faster training in distributed classifier approach compared to single machine processing. With big dataset processing in non distributed environment was only possible as processing is done on block basis. If block size is increased to reach around 4 blocks of 10,000 samples each, it is not possible to process on a single machine using support vector machines ,which is an inherent limitation of the method. So for the purpose of generalization for bigger block sizes in cloud, we evaluate using distributed classifier. In our experiments, we considered three different categories of covert timing channels based on the resources exploited. In replicating the attack ,we take utmost care to make sure the stealth of the attack is high. We build signature using time interval, k-s test score and context variables. The signature of applications' resource behavior now goes down the pipeline and classifier module processes it. Use of support vector machine based classifier is important module because of the following reason. It is depicted in the graph 1, k-s test scores of time interval in covert and overt traffic, show very close scores and overlap in certain cases making this feature unreliable(shown with arrows). It is visible that after certain time intervals bus channel and cache based attacks show overlap . This will be difficult to detect if we merely rely on k-s test scores. Thus use of context and machine learning will facilitate detection.

### 4.1 CPU load based attacks

In CPU-load based attack[19-20], attacker VM launches an attack by making sudden bursts of requests for CPU resource. For a covert channel attack sender sends more than 100 user requests in burst to transmit a bit 1 and no request is made for a bit 0.With time synchronization after a time gap of every 10 secs request for the resource is made. One of the colluding VM with sensitive information intending to leak it, has Apache JMeter tool, usually used to load test a website and spy VM has a static website running on it. We can use JMeter to specify the user load and control cpu usage of the system which gives spy 1 bit if load is high and 0 bit if load is low. Also the attack can be used to launch denial of service against a VM with web server by sending request bursts with 500 or 1000 users (threads) as it takes CPU usage upto 99%. A simplified version exploited CPU as shared resource on a windows based system. Monitoring tool can be used by the receiver to interpret the bits. In covert channel attacks care has to be taken not to make time difference very significant using load as covertness is inversely proportional to accuracy and bandwidth of the channel making it easily detectable, so based on the system we can decide ideal number of threads as we run the application. Depicted in Figure 3 are corresponding variations in resource monitor. The windows based setup is used along with debian to ensure attack parameters confirm to standards in different platforms. Tools Used:Jmeter(java desktop application designed to load test behaviour and performance);Static web site(created for experiment).; 2 Virtual Machines each with following configuration: 512 MB RAM (expandable),1 cpu core,127 GB HDD.

|  | N= 5000 |  | N= 200 |  |
| --- | --- | --- | --- | --- |
|  | Overt | Covert | Overt | Covert |
| Overt | 96 | 4 | 86 | 14 |
| Covert | 0 | 100 | 0 | 100 |

**Table 2: Confusion matrix when testing presence of CPU load based channels(in %)**

### 4.2 Cache based attacks

Cache based covert channels don't have a very high bandwidth or accuracy as cache scheduling uncertainty in both lab and Amazon EC2 leads to error rate of around 30 percent to 85 percent [22] while raw bandwidth is around 190 kb/s and though cache based covert channels have a number of limitations in virtual environment like migration etc nonetheless it offers a good case to be studied for our detection framework. In [21] authors have experimentally shown access latency caused to attack. As in cloud access time is the resource being exploited, so receiver first tests access latency by loading a file as big as cache size of the machine to fill entire cache and then access it to keep track of lower threshold of access latency. Then in a 2 VM setup whenever sender VM wants to send bit 1 then it fills cache will equally large file to fill cache so that when at stipulated time interval receiver VM access its file it will cause higher latency than measured earlier. This way attacker(sender) and spy(receiver) synchronize to leak sensitive bits of data and the process continues till all the bits are leaked.

|  | N= 5000 |  | N= 200 |  |
| --- | --- | --- | --- | --- |
|  | Overt | Covert | Overt | Covert |
| Overt | 90 | 10 | 80 | 20 |
| Covert | 0 | 100 | 0 | 100 |

**Table 3: Confusion matrix when testing presence of cache based channels(in %)**

### 4.3 Memory bus based attacks

Whenever memory bus is put in contended state as its being shared by VMs and there is a delay in accessing the bus,it will be noticed by the receiver VM. This attack as explained in [21] is carried out by a sender sending the packet by causing latency in accessing the memory bus by the receiver. This is interpreted as high and low signal. Wu et al have demonstrated that bus contention based attack can achieve a high bandwidth of over 100-bits per second which can steal a thousand byte private key file in less than 3 minutes. This attack is hard to detect because of its negligible impact on performance of cache causing transparency for any cache based detection techniques. This validates our choice of parameters .It will be hard to capture such attacks by performance based detection frameworks.

Minimum latency occurs when transmitting a 0. Latency is calculated as Mean(Access Time)- Base time. In bus contention based channel ,contention is created on front side bus using atomic instructions xchg 8086 instruction for exchanging memory with register which forcefully locks bus.There are many instructions in assembly that can be used to deliberately create contention but might require use of lock prefix. A predefined threshold time is agreed between two programs. Then if at receiver end mean(access time) > threshold bit is interpreted as 1 else 0. If too many consecutive 1s or 0s,receiver can assume that sender process has been rescheduled by the hypervisor.

|  | N= 5000 |  | N= 200 |  |
| --- | --- | --- | --- | --- |
|  | Overt | Covert | Overt | Covert |
| Overt | 94 | 6 | 84 | 16 |
| Covert | 0.5 | 99.5 | 4 | 96 |

**Table 4 : Confusion matrix when testing presence of memory bus based channels(in %)**

### 4.4 Performance metrics

We used MATLAB's built in SVM-classifier initially and then we distributed the code over different VMs for processing. The algorithm 4 and 5 shown in section 3.3.2 is implemented in Matlab and cross validated across machines.

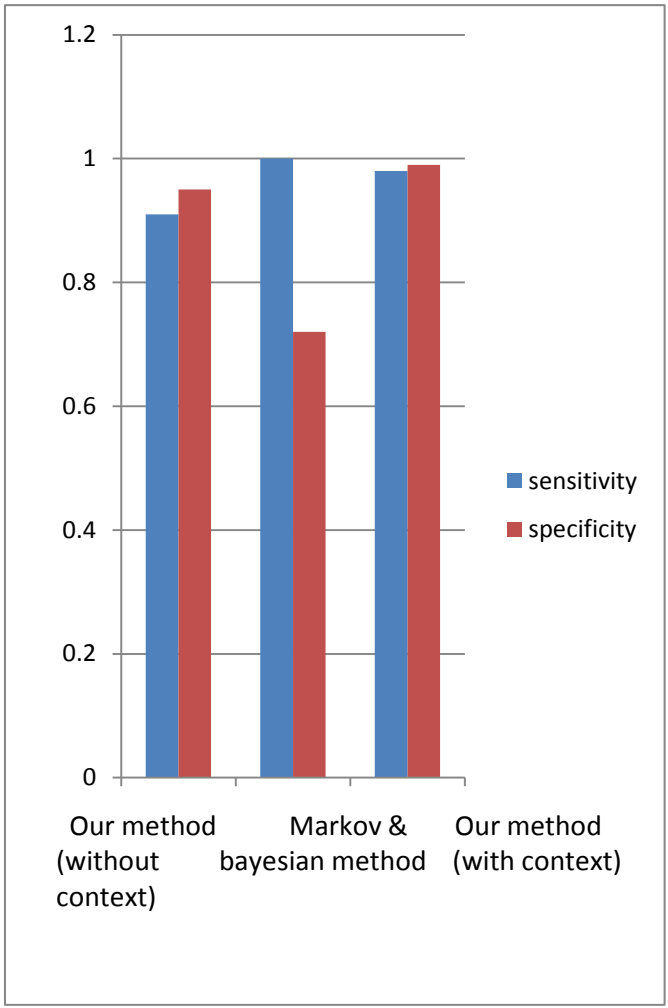

**Figure4: graph showing specificity and sensitivity**

The classifier was optimized using Box –Constraint and Kernel parameter was set to rbf. The non linear transform using Radial Basis function is able to provide separation even when circular line is required for separating in x-space(non- hyperplane space). The parameters worked well even when we input certain non separable data. Though to reduce training and cross validation error but at the same time we did not overfit the function as we used distributed instances of SVM and threshold values were different so we took an average of values to calculate one acceptable threshold .Arguably authors believe average best depicts fits while making choice out of multiple instances of same variable. Context variables have a significant contribution which becomes apparent from our experimental results. To verify we trained and tested with and without the use of context. As shown in figure 4, the gap between specificity and sensitivity closes with the use of context. Besides our approach as compared to $C^2$ detector performs well.

In absence of real life cloud which is a noisy environment , authors check the robustness of method proposed and how fast it would deteriorate in performance if we added some noise to the data. In fact in practical scenarios such situations will be common.5% gaussian noise is added to vector latency. The error rate increased by .6% which is under tolerance limit .Also for comparison purposes and for the sake of completeness in validating our approach in comparison with contemporary approaches, in our tests we applied Neural network for training and testing on data using 10-fold cross validation. The results are shown in Graph3.The choice of SVM as a classifier had many advantages over ANN like ability to converge faster, accuracy. By theory SVM is less prone to overfitting and has global unique maxima unlike ANNs. We tested NN performance and error rate was higher at 8% in first training. It did improve as we retrain the network but as our model is required for real time scenarios, time required for training should be minimum. Initially we trained NN for three features on 6 neurons(taking ideal case of 2N+1). Thus comparative analysis yields better results with SVM.

## 5. Summary and Conclusion

Current techniques of covert channel detection target very specific covert channels which means they wont be able to detect channels which are related. Hence we have proposed a method which doesn't lose original parameters that will help in extension of our method and enable generalization.
In the paper we have used performance metrics and high activity as parameters for detection of the bus contention time channel but as we try to generalize this model, we might come across time channels that are clever enough to leave no footprints, keep activity levels lower than the threshold and do not affect the performance of the system. Therefore in future we plan to propose a detection framework that is more robust to such techniques and and generalizable to build signatures or profiles of new anomalies in real time. In future we plan to propose a framework for anomaly detection in cloud ,which is highly dynamic and the approach should be able

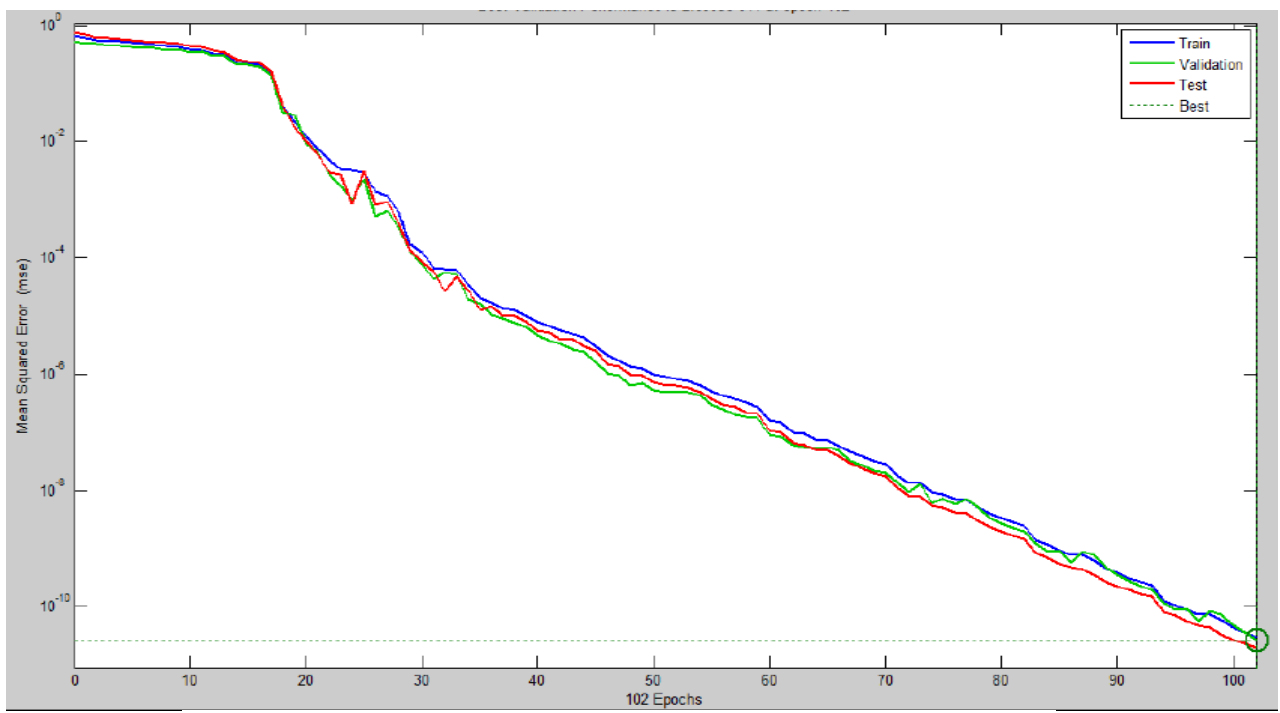

**Figure 5 : graph with ANN as a classifier (limitations)**

to cover all possible parameters so that adaption is possible and new anomalies can be covered.

## Appendix A. Screenshots and Graphs

The framework has been realized based on experiments conducted for data collection. The graphical tools and C language, matlab codes were used for data collection. Tools used for conducting experiments are shown in following subsections along with snapshots.

### A.1 Screenshots

Jmeter is a pure java based graphical server performance testing tool, for both static and dynamic resources (files or CGI, Servlets, Perl scripts). It is designed to test load, functional performance, regression etc. The screenshot shows we dramatically increased load on server which contains our static website. The performance monitor captures results as shown in figure B.

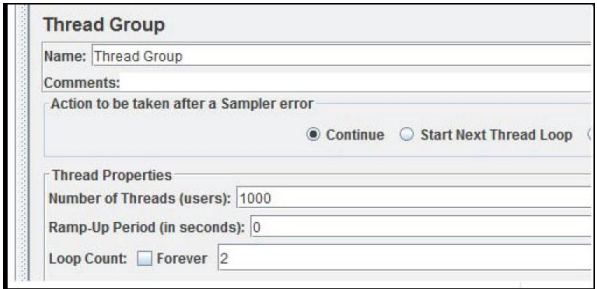
Fig-A: Snapshot of JMeter Tool

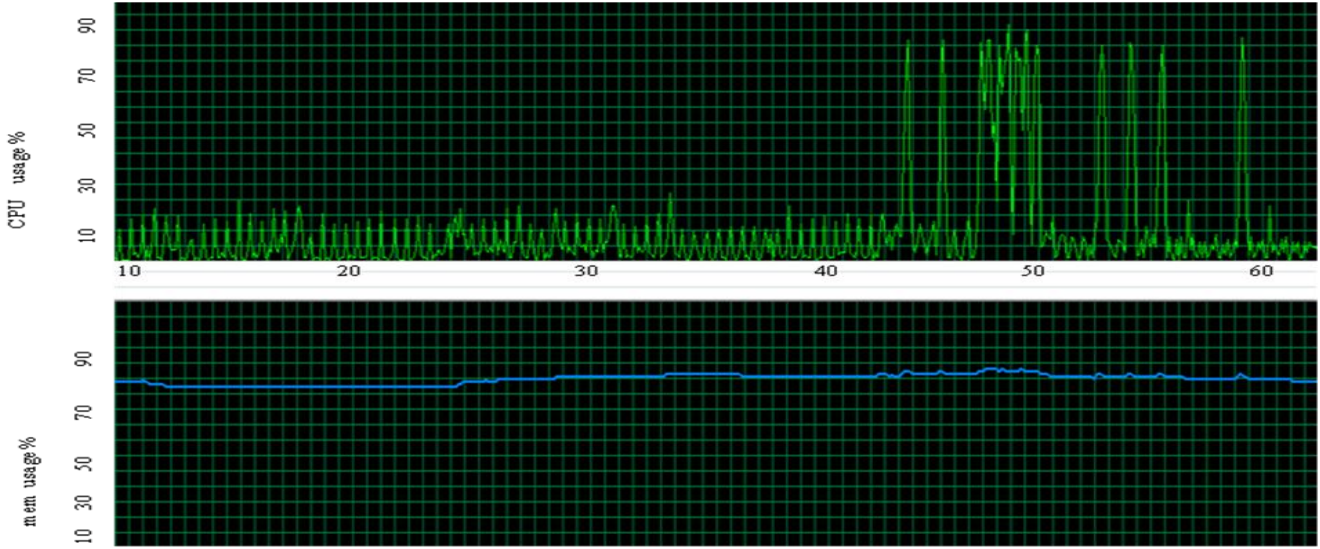

**Figure B: Graph showing resource usage changes**

### A.2 Tool matlab

To enable noisy environment in our cloud testbed, we used matlab functionality. The graph in figure C depicts SNR or signal to noise ratio between the actual data and the syntheticall added noise. This graph is important to show that the noise was distributed throughout.

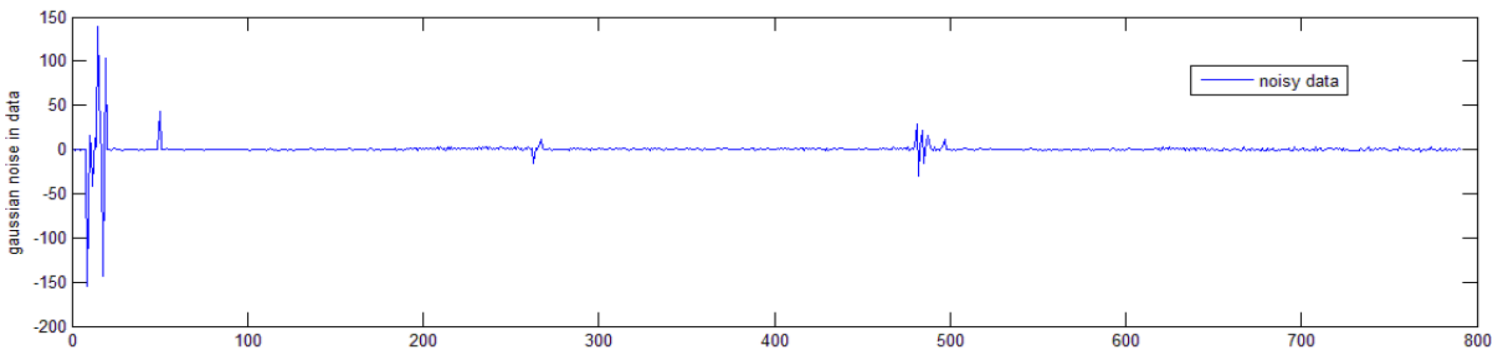

**Figure C: SNR in matlab**